\documentstyle[11pt,epsfig,twocolumn]{article}
\newcommand{\beq}{\begin{equation}}
\newcommand{\eeq}{\end{equation}}
\newcommand{\beqa}{\begin{eqnarray}}
\newcommand{\eeqa}{\end{eqnarray}}
\newcommand{\ba}{\begin{array}}
\newcommand{\ea}{\end{array}}

\begin{document}
    \markright{\rm \hfill M. Robnik, L. Salasnich and M. Vrani\v car:
    WKB corrections to the energy splitting ...\hfill}
    \pagestyle{myheadings}
    \thispagestyle{plain}
\twocolumn[\hsize\textwidth\columnwidth\hsize\csname %
@twocolumnfalse\endcsname
  \begin{center}

  \noindent Submited to {\it Nonlinear Phenomena in Complex Systems}
  \hfill
   Preprint CAMTP/98-4
   \end{center}

\begin{center}
{\LARGE \bf WKB corrections to the energy splitting 
\\[0.15cm]
in double well potentials}\\
\vspace{0.4cm}

{\large
Marko Robnik$^\dag$,
Luca Salasnich$^\ddag$ and
Marko Vrani\v car$^\dag$}
\vspace{0.15cm}

{\small \it \dag Center for Applied Mathematics and Theoretical Physics,
 University of Maribor,\\ 
 Krekova 2, SI-2000 Maribor, Slovenia\\
 \ddag Istituto Nazionale per la Fisica della Materia, 
                Unit\`a di Milano, \\
 Dipartimento di Fisica, Universit\`a di Milano, \\
 Via Celoria 16, I-20133 Milano, Italy \\[0.15cm]
 E-mail: robnik@uni-mb.si, 
         salasnich@mi.infm.it and 
         mark.vranicar@uni-mb.si \\
{\rm\small (Submited 7 June  1999)}
}
\end{center}
\parindent 0pt
\parskip 4pt

\begin{quotation}{
By using the WKB quantization 
we deduce an analytical formula 
for the energy splitting in a double--well potential 
which is the usual Landau formula 
with additional quantum corrections. 
Then we analyze the accuracy of our formula for 
the double square well potential, the inverted harmonic 
oscillator and the quartic potential. 
\parskip 10pt
\parindent 0pt

\vspace{0.5cm}
}
\end{quotation}
]

\section{Introduction}
Semiclassical quantization is very useful to understand the global 
behaviour of eigenfunctions and energy spectra of quantum systems, 
since it allows us to obtain analytic expressions. 
The leading semiclassical approximation 
(torus quantization) is just the first term of a certain 
$\hbar$--expansion, which is called WKB (Maslov and Fedoriuk 1981). 
A systematic study of the accuracy 
of semiclassical approximation is very 
important, especially in the context of quantum chaos (Casati and Chirikov 
1995, Gutzwiller 1990). Since this is a difficult task, it has been 
attempted for simple systems, where in a few cases 
even exact solutions may be worked out (Bender, Olaussen and Wang 
1977, Voros 1993, Robnik and Salasnich 1997a,b, Salasnich and Sattin 1997). 

In this paper we analyze the energy splitting 
of doublets in a generic one--dimensional double--well potential. 
By using the WKB quantization we deduce an analytical formula 
for the energy splitting which is the usual Landau (1997) formula 
with additional quantum corrections. The splitting 
formula can be formally written as 
\beq 
\Delta E = A \exp{\Big[-{S\over \hbar }\Big]} \, ,
\eeq
where $S$ is the usual classical action 
inside the classically forbidden region (between the two 
turning points) 
and $A$ is called the tunneling amplitude, which can be written as a 
polynomial expression, expanded in powers of $\hbar$. 
This formula is based on a linear approximation of the 
potential near the turning points. First 
we introduce the basic definitions, then 
we derive in detail the splitting formula by using the semiclassical 
WKB expansion and finally we study its  
validity for the double square well potential,  
the inverted harmonic oscillator and the quartic potential. 
Our present work is essentially a semiclassical expansion theory
of the tunneling amplitude $A$ of equation  (1). We shall
demonstrate that it is indeed a significant improvement
of the Landau approximation  (1997). Also, we shall
mention some potential applications.

\section{Basic formalism}
Let us consider a one--dimensional system with Hamiltonian
\beq
H = {p^2\over 2m} + V(x) \, ,
\eeq 
where $V(-x)=V(x)$ is a symmetric double--well potential. 
The stationary Schr\"odinger equation of the system reads 
\beq
{\hat H} \psi (x) = \Big( -{\hbar^2\over 2m} {d^2 \over dx^2} + V(x) \Big) 
\psi (x) = E \psi (x) \, . 
\eeq
The Sturm--Liouville theorem (see, for example, Courant and Hilbert
1968) ensures that for one--dimensional systems 
there are no degeneracies in the spectrum. Let $\psi_1$ and $\psi_2$ be 
two exact eigenfunctions of the Schr\"odinger equation 
\beq
{\hat H} \psi_1 = E_1 \psi_1 \;\;\;\;\; \hbox{and} \;\;\;\;\; 
{\hat H} \psi_2 = E_2 \psi_2 \, ,
\eeq
such that $\psi_1(-x)=\psi_1(x)$ and $\psi_2(-x)=-\psi_2(x)$ 
and $E_1\simeq E_2$. To calculate the splitting 
$\Delta E = E_2 - E_1$, we multiply the first equation by $\psi_2$ 
and the second by $\psi_1$ and then we subtract the two resulting 
equations. By integrating from $0$ to $\infty$ we find
\beq
\Delta E = {\hbar^2 \over 2 m} 
{ \psi_1(0)\psi{'}_2(0) - \psi{'}_1(0)\psi_2(0) \over 
\int_0^{\infty} \psi_1(x) \psi_2(x) dx } \, .
\eeq
We write the eigenfunctions $\psi_1$ and $\psi_2$ in terms 
of the right--localized function 
\beq
\psi_0(x) = {1\over \sqrt{2}}(\psi_1(x) + \psi_2(x) )\, .  
\eeq
It is easy to show that $E_0 = <\psi_0 | {\hat H} |\psi_0 > 
= {1\over 2} (E_1 + E_2)$. Then, with the approximation
$\int_{0}^{\infty} \psi_0^2dx \approx 1$, committing an exponentially
small error, namely
\beq 
\int_0^{\infty} \psi_1(x) \psi_2(x) dx = 
\int_0^{\infty} \psi_0^2(x) dx - {1\over 2} \approx {1\over 2} \, ,
\eeq
we get 
\beq
\Delta E = {2\hbar^2 \over m} \psi_0(0)\psi{'}_0(0) \, ,
\eeq
which is an almost exact starting 
formula to calculate the energy splitting, since 
the error committed in approximation (7) is exponentially small. 
One should observe that this quantity is always positive,
because the tail of the right localized eigenfunction
$\psi_0(x)$ at $x=0$ has the same sign for $\psi_0(0)$
and its derivative $\psi'_0(0)$. Another way to see this is
to realize that due to the Sturm-Liouville theorem there
are no degeneracies in one--dimensional systems, implying that all pairs of
almost degenerate states, from the ground state up, 
are grouped by odd state above the even state.

\section{Semiclassical method}
To determine the function $\psi_0$ we perform a WKB expansion 
of the Schr\"odinger equation. We observe that 
a generic eigenfunction $\psi$ of the Schr\"odinger equation 
can always be written as
\beq
\psi (x) = \exp{ \big( {i\over \hbar} \sigma (x) \big) } \, ,
\eeq
where the phase $\sigma (x)$ is a complex function that satisfies 
the Riccati differential equation
\beq
\sigma{'}^2(x) + ({\hbar \over i}) \sigma{''}(x) = 2m(E - V(x)) \, .
\eeq
The WKB expansion for the phase is given by
\beq
\sigma (x) = \sum_{k=0}^{\infty} ({\hbar \over i})^k \sigma_k(x) \, .
\eeq 
Substituting (11) into (10) and comparing like powers of $\hbar$ gives 
the recursion relation ($n>0$) (see Bender, Olaussen and Wang 1977) 
\beq
\sigma{'}_0^2=2m(E-V(x)) \, , \;\;\;\; 
\sum_{k=0}^{n} \sigma{'}_k\sigma{'}_{n-k}
+ \sigma{''}_{n-1}= 0 \, .
\eeq
With the momentum $p=\sqrt{2m(E-V(x))}$ the first five orders in the 
WKB expansion are given by 
\beqa
\sigma'_0 & = & p\, , \nonumber \\[3mm]
\sigma'_1 & = & -{p{'}\over 2p} \, , \nonumber \\[3mm]
\sigma'_2 & = & {p{''}\over 4 p^2} -{3\over 8}{p{'}^2\over p^3} \, , 
                \nonumber\\[3mm]
\sigma'_3 & = & \frac{p'''}{8p^3}+\frac{3}{4}\frac{p'p''}{p^4}-
             \frac{3}{4}\frac{p^{'3}}{p^5} \, , \nonumber \\[3mm]
\sigma'_4 & = & \frac{1}{16}\left(
             \frac{p''''}{p^4}-10\frac{p'''p'}{p^5}
             -\frac{13}{2}\frac{p^{''2}}{p^5}
             \right) \nonumber \\[2mm]
           & & +\frac{1}{16}\left( \frac{99}{2}\frac{p''p^{'2}}{p^6}
             -\frac{297}{8}\frac{p^{'4}}{p^7}\right) \, , \\[3mm]
\sigma'_5 & = & \frac{1}{32} \left(
             -\frac{p'''''}{p^5}+15\frac{p''''p'}{p^6}
             +24\frac{p'''p''}{p^6}\right) \nonumber \\[2mm]
           & &+\frac{1}{32}\left(-111\frac{p'''p^{'2}}{p^7}
              -144\frac{p^{''2}p'}{p^7}\right) \nonumber \\[2mm]
           & &+\frac{1}{32}\left(
             510 \frac{p''p^{'3}}{p^8}
            -306\frac{p^{'5}}{p^9}\right) \nonumber \, .
\eeqa
In particular, if we call $a$ and $b$ the two turning points 
corresponding to the energy $E$, 
the right localized wavefunction $\psi_0$ is given by 
$$
\psi_{0}(x)=\frac{C_1}{\sqrt{|p|}}\exp\left[\frac{i}{\hbar}
                  \left(\int_a^x |p|\,dx
                  +\sigma_{even}\right)+\frac{1}{\hbar}\sigma_{odd}\right] 
$$
\beq              
+\frac{C_2}{\sqrt{|p|}}\exp\left[-\frac{i}{\hbar}
                   \left(\int_a^x |p|\,dx+\sigma_{even}\right)
                   +\frac{1}{\hbar}\sigma_{odd}\right] \, , 
\eeq
for $a<x<b$ (allowed region), where 
$$
\sigma_{even}=\sum_{k=1}^{\infty}(-1)^k\hbar^{2k}\sigma_{2k}(|p(x)|) 
$$
\beq  
{\rm with} \;\;
\sigma_{2k}(-|p|)=-\sigma_{2k}(|p|)\, ,
\eeq
and
$$
\sigma_{odd}=\sum_{k=1}^{\infty}(-1)^k\hbar^{2k+1}\sigma_{2k+1}(|p(x)|) 
$$
\beq
{\rm with}\;\; 
\sigma_{2k+1}(-|p|)=\sigma_{2k+1}(|p|)\, . 
\eeq
Instead we get
\beq
\psi_{0}(x) = 
\frac{C_a}{\sqrt{|\tilde{p}|}}\exp\left[\frac{1}{\hbar}\left(\int_a^x
|\tilde{p}|\,dx+\tilde{\sigma}_{even}+\tilde{\sigma}_{odd}\right)\right] \, ,
\eeq
for $x<a$ (forbidden region), and also
\beq
\psi_{0}(x)
=\frac{C_b}{\sqrt{|\tilde{p}|}}\exp\left[\frac{1}{\hbar}
\left(-\int_b^x |\tilde{p}|\,dx-\tilde{\sigma}_{even}
+\tilde{\sigma}_{odd}\right)\right] \, ,
\eeq
for $b<x$ (forbidden region), where 
$$
\tilde{\sigma}_{even}=\sum_{k=1}^{\infty}\hbar^{2k}\tilde{\sigma}_{2k}
(|\tilde{p}(x)|) \;\;
$$
\beq
{\rm with}\;\; 
\tilde{\sigma}_{2k}(-|\tilde{p}|)=-\tilde{\sigma}_{2k}(|\tilde{p}|)
\, ,
\eeq
and
$$
\tilde{\sigma}_{odd}=\sum_{k=1}^{\infty}\hbar^{2k+1}\tilde{\sigma}_{2k+1}
(|\tilde{p}(x)|) 
$$
\beq
{\rm with}\;\;
 \tilde{\sigma}_{2k+1}(-|\tilde{p}|)=\tilde{\sigma}_{2k+1}
(|\tilde{p}|)\, ,
\eeq
where $\tilde p=\sqrt{2m(V(x)-E)}$ and 
$\tilde \sigma'_k(x)=\sigma'_k(\tilde p(x))$.
In evaluating the integrals $\sigma_k$ using (13) we cannot
integrate naively on the real axis, because such integrals
are divergent, but must take a partial derivative w.r.t.
the energy of certain fundamental complex contour integral.
See (Bender {\em et al} 1977) and section 6.

We observe that it is possible to write $C_a$, $C_b$, $C_1$ and $C_2$ 
in terms of a unique parameter $C$ by imposing 
the uniqueness of the wavefunction $\psi_0$ at the turning points. 
Following Landau (1997) and Merzbacher (1970) 
we suppose that near the turning point $x=a$ 
it is possible to approximate the potential locally linearly by writing 
\beq
E - V(x) = F_0 (x-a) \, ,
\eeq
with $F_0 > 0$. In this case the connections at the turning point 
imply that 
\beq
C_a=C_b=C \; , \;\; C_1 = C e^{i {\pi \over 4}}  \;, \;\;  
C_2 = C e^{-i {\pi \over 4}} \, ,
\eeq
and the right localized function $\psi_0(x)$ can be written 
for $|x|<a$ as 
\beq
\psi_0 (x)=\frac{C}{\sqrt{|\tilde{p}|}}
\exp\left[\frac{1}{\hbar}\left(\int_a^x
|\tilde{p}|\,dx+\tilde{\sigma}_{even}+\tilde{\sigma}_{odd}\right)\right] \, .
\eeq
\newpage
In this way the splitting formula, up to the 5th order, after taking into
account some straightforward relations for $\sigma_k$ and its
derivatives (see the Appendix),  becomes
\beqa
\Delta E & = & \frac{2\hbar C^2}{m} \left[\frac{}{}\right. 
                1+\hbar 2\tilde{\sigma}_2(0)
              +\hbar^2 2\tilde{\sigma}^2_2(0)  \nonumber \\[2mm]
         &   & +\hbar^3
               \left(2\tilde{\sigma}_4(0)+\frac{4}{3}\tilde{\sigma}^3_2(0)
               \right) \nonumber \\[2mm] 
         &   & \left.
               +\hbar^4 \left( \frac{2}{3}\tilde{\sigma}_2^4(0)
               +4\tilde{\sigma}_2(0)\tilde{\sigma}_4(0)\right)\right] 
                \\[2mm]
         &   & \times \exp\left[-\frac{2}{\hbar}\int_0^a|\tilde{p}|\,dx\right] 
               \, . \nonumber
\eeqa
The equation (24) can be written in a compact form,
certainly up to the 5th order, and probably also generally
to all orders,  as follows
\beq
\Delta E=\frac{2\hbar C^2}{m}\exp\left[\frac{2}{\hbar}
\left(-\int^a_0|\tilde p|\,dx+\tilde \sigma_{even}(0)\right)\right] \, ;
\eeq
$$\tilde \sigma_{even}(0)=\int_a^0 \tilde 
\sigma'_{even}(|\tilde p(x)|)\, dx\, .
$$
To determine $C$ we impose the normalization condition 
\beq
1 = \int_{0}^{\infty} |\psi_0(x)|^2 dx \, ,
\eeq
from which we get 
\beq
2 C^2 \int_a^b \; {1\over p} 
\exp{\big(2{\sigma_{odd}\over \hbar}\big)} dx = 1 \, , 
\eeq
and an expression for $C^2$,
\beqa
C^2&=&\frac{1}{2}\left[\int_a^b\frac{1}{p}\,dx\right]^{-1}
\left\{1+2\hbar^2\frac{\int_a^b\frac{\sigma_3}{p}\,dx}{\int_a^b\frac{1}{p}\,dx}
\right. \nonumber \\[2mm]
& &\left.
+2\hbar^4\left(2\left[\frac{\int_a^b\frac{\sigma_3}{p}\,dx}{\int_a^b\frac{1}{p}\,dx}\right]^2
-\frac{\int_a^b\frac{\sigma^2_3+\sigma_5}{p}\,dx}{\int_a^b\frac{1}{p}\,dx}
\right)\right\}\, , \nonumber \\
\eeqa
which is valid up to the 4th order in the $\hbar$ power series
of $C^2$.
The final formula is given by \\
\beqa
\Delta E &  = & \frac{\hbar}{m}\left[\int_a^b\frac{1}{p}\,dx\right]^{-1}  
                \nonumber\\[2mm]
         & \times &  \left\{\frac{}{}1+2\hbar\tilde{\sigma}_2(0)
                +2\hbar^2\left(\tilde{\sigma}_2^2(0)+
                \frac{\int_a^b\frac{\sigma_3}{p}\,dx}
                 {\int_a^b\frac{1}{p}\,dx}\right)\right.  
                \nonumber  \\[2mm]
         &   &  +2\hbar^3\left(\tilde{\sigma}_4(0)+
               \frac{2}{3}\tilde{\sigma}_2^3(0)+2\tilde{\sigma}_2(0)
               \frac{\int_a^b\frac{\sigma_3}{p}\,dx}
               {\int_a^b\frac{1}{p}\,dx}\right)       \nonumber  \\[2mm]
         &   &  +2\hbar^4\left(\frac{1}{3}\tilde{\sigma}_2^4(0)+
                2\tilde{\sigma}_2(0)\tilde{\sigma}_4(0)\right) 
                \nonumber \\[2mm]
         &   &  +2\hbar^4\left(2\tilde{\sigma}^2_2(0)
                \left[\frac{\int_a^b\frac{\sigma_3}{p}
               \,dx}{\int_a^b\frac{1}{p}\,dx}\right]\right) \\[2mm]
         &   & 
              +2\hbar^4\left.\left(2\left[\frac{\int_a^b\frac{\sigma_3}{p}\,dx}
            {\int_a^b\frac{1}{p}\,dx}\right]^2-
            \frac{\int_a^b\frac{\sigma^2_3+\sigma_5}{p}\,dx}
            {\int_a^b\frac{1}{p}\,dx}\right)\right\} \nonumber \\[2mm]
        &    & \times\exp\left[-\frac{2}{\hbar}\int_0^a|\tilde{p}|\,dx\right] 
               \, . 
               \nonumber
\eeqa 
This formula is the usual Landau (1997) formula for the energy splitting 
(1st order in $\hbar$ for the tunneling amplitude) 
with additional quantum corrections (up to the 5th order 
in $\hbar$ for the tunneling amplitude). We note that 
higher--order WKB corrections quickly increase in complexity 
(Robnik and Salasnich 1997a,b) but, in principle, 
they can be calculated from the equation (12). 
It is important to stress that 
our splitting formula is good if the potential is sufficiently smooth 
so that the linear approximation is valid near the turning points. 

The same splitting formula  (25) and (29)  can be derived using the
semiclassical scattering formalism, for example as expounded
by Iyer and Will (1987) and by  Will and Guinn (1988).
However, the main result (25) and (29) of this paper cannot be
obtained by simple substitution or reinterpretation of their
results. They use the scattering approach and treat the
scattering problem (asymptotical free motion), calculating
the transmission coefficients for the tunneling penetration
through a potential barrier, and specifically they treat
the behaviour near to the top of the potential barrier.
Also, their approach is doubly perturbative, namely they
make the power expansion of the potential around the top of 
the barrier whose leading term is of course the inverted
harmonic oscillator plus power terms in the series expansion,
and they do at the same time the semiclassical expansion
in terms of the powers of the Planck constant $\hbar$.
They offer a formalism (an algorithm) how to calculate 
the requested quantities (transmission coefficients) to all 
orders, however they solve the relevant equation only up to the fourth
order. So, their result cannot be easily mapped (by
substitutions and other simple operations) onto
our problem and our solution. Indeed, to get the
result (25) and (29) using the scattering approach it is necessary
to go back to the very first step in their formalism. 

We have done this and confirmed, as mentioned above, that the
result is the same. To this end the solution is written down
in the form of the semiclassical ansatz in each of five
regions separated by the turning points. At the turning points
we do not request the condition of the continuity of the
wavefunction and its derivative, but use the so-called
Kramers correspondence rules instead (they determine the 
coefficients of the ansatz in such a way, that exponentially
increasing solution in the classically forbidden regions does 
not occur). Thus, the asymptotic boundary conditions for 
the scattering problem are automatically taken into account:
no propagation to the left or to the right of the classically
allowed region.

Using this ansatz we have six unknown
coefficients plus the eigenenergy that we seek, and we have
six linear equations plus the normalization condition,
so the problem is well defined.
This system of equations can be reduced by simple elimination of
some of the coefficients to a set of two homogeneous
linear equations for which the solvability condition is
now vanishing of its determinant, which must be satisfied
precisely at the eigenvalues of the energy. Assuming that
the pair of almost degenerate levels is separated by a small
amount $\Delta E$, we can do the Taylor expansion up to
the first order in $\Delta E$, neglect the quadratic and higher
terms, and obtain exactly the equation (25) and (29).

\section{Double square well potential}
As the first example, we consider the double square well potential. 
In this case the linear approximation of the potential near the turning 
point is not valid. The potential is given by  
\beq
V(x)=\left\{ 
\ba{cc}
V_0    & \hbox{ for } |x| < a  \\
0      & \hbox{ for } a < |x| < b \\
\infty & \hbox{ for } |x| > b 
\ea
\right. 
\eeq
For this potential we have $p{'}(x)=p{''}(x)=0$ for $-a<x<a$ and 
the corrections to the Landau (1997) formula are zero. 
A naive application of the splitting formula gives 
\beq
\Delta E = {2\hbar \sqrt{E} \over \sqrt{2m} (b-a)} 
\exp{\Big(- {2a\over \hbar} \sqrt{2m(V_0-E)} \Big)} \, .
\eeq 
This formula is {\it not} correct. In fact, by using 
the exact\footnote{Actually, strictly speaking, this is not 
exact but nevertheless the same expression that we get 
by evaluating the leading term for $\Delta E$ by using the 
implicit trigonometric eigenvalue equation 
(Fl\"ugge 1971, Robnik and Salasnich 1997, unpublished).} 
wavefunction
\beq
\psi_0(x) = D \exp{\Big(-{1\over \hbar}\sqrt{2m(V_0-E)}x \Big)} 
\eeq
for $0<x<a$ (forbidden region), and 
\beq
\psi_0(x) = A \exp{\Big({i\over \hbar}\sqrt{2mE}x \Big) } + 
B \exp{\Big(-{i\over \hbar}\sqrt{2mE}x \Big) } 
\eeq
for $a<x<b$ (allowed region), and by imposing the exact matching 
and normalization conditions (Fl\"ugge 1971) we find
\beqa
A& =& {D\over 2} \left( 1 - i \sqrt{V_0-E\over E} \right)\nonumber \\[2mm]
&& 
\times\exp{\Big( {a\over \hbar} 
\sqrt{2m(V_0-E)} - {a\over \hbar}\sqrt{E} \Big)} \, , \nonumber \\
\eeqa
\beqa
B& =& {D\over 2} \left( 1 + i \sqrt{V_0-E\over E} \right) \nonumber \\[2mm]
&& 
\times\exp{\Big( {a\over \hbar} 
\sqrt{2m(V_0-E)} + {a\over \hbar}\sqrt{E} \Big)} \, , \nonumber \\
\eeqa
and 
\beq
D^2= {2E\over V_0 (b-a)} \nonumber 
\exp{\Big( - {2a\over \hbar} \sqrt{2m(V_0-E)} \Big)} \, .
\eeq
Then we obtain: 
\beqa
\Delta E &= &{4 \hbar E \sqrt{2m(V_0-E)} \over m V_0 (b-a)} \nonumber \\[2mm]
&&
\times\exp{\Big( -{2a\over \hbar} \sqrt{2m (V_0 -E)} \Big) } \, .
\eeqa
This is the exact energy splitting for the double square well potential. 
It differs by a factor $4\sqrt{E(V_0-E)}/V_0$ from the WKB result 
based on the connection formulae (21--22) which are not justified 
in the present case. 

\section{Inverted harmonic oscillator}
In this section we compare the Landau formula of the energy splitting 
with the exact one. We consider the inverted harmonic oscillator given by
\beq
V(x)=\left\{ 
\ba{cc}
V_0 \big(1-{x^2\over a^2}\big) & \hbox{ for } |x| < a  \\
0         & \hbox{ for } a < |x| < b   \\
\infty    & \hbox{ for } |x| > b   \\
\ea
\right. 
\eeq 
We can introduce the following reduced variables
\beq
\bar{x}={x\over a}\; , \;\;\; \bar{b}={b\over a}\; ,\;\;\; 
\bar{E}={E\over V_0}\; , \;\;\; 
\hbar_{eff}={\hbar \over a\sqrt{mV_0}} \, . 
\eeq
Then it is not difficult to show that the Landau formula reads  
\beq
\Delta \bar{E} = {2\hbar_{eff} D^2} 
\exp{\Big( -{\pi \over \hbar_{eff}\sqrt{2}} (1-\bar{E}) \Big) } \, , 
\eeq
where 
\beq
D^2 = 
\Big[{1\over 2}\ln{ \Big( {1+\sqrt{\bar{E}}\over 1-\sqrt{\bar{E}} } \Big) }  
+ {(\bar{b}-1)\over \sqrt{\bar{E}} }  \Big]^{-1} \, .  
\eeq
This formula can be compared with the exact energy splitting.  
Let $\psi (x)$ be a quantum state of the inverted harmonic oscillator. 
It can be written in terms of the eigenstates $\psi^{(+)}$ and $\psi^{(-)}$ 
of the square well potential
\beq
\psi(x)=\sum_{n=1}^{\infty}(a_n\psi_n^{(+)}+b_n\psi_n^{(-)}) , ,
\eeq
where 
$$
\psi_n^{(+)}=\frac{1}{\sqrt{b}}\cos\left[\pi\frac{2n-1}{2b}x\right] \, ,
$$
$$
\psi_n^{(-)}=\frac{1}{\sqrt{b}}\sin\left[\pi\frac{n}{b}x\right] \, .
$$
The matrix elements of the quantum Hamiltonian of the inverted 
harmonic oscillator read
$$
H^{(+)}_{m,n}=\int_{-b}^{b}\psi_m^{(+)}{\hat H}\psi_n^{(+)}\,dx =
$$
\beq 
=\left\{
\begin{array}{ll}
m=n:
& \frac{\hbar^2_{ef}}{2}\left(\frac{2n-1}{2b}\pi\right)^2+\frac{2}{3b} \\[2mm]
&      +\frac{2b^2}{\pi^3(2n-1)^3}\sin\left[\frac{2n-1}{b}\pi\right] \\[2mm]
&      -\frac{2b}{(2n-1)^2\pi^2}\cos\left[\frac{2n-1}{b}\pi\right]\\[5mm]
m\neq n:
&\frac{2b^2}{\pi^3}\left(\frac{\sin\left[\frac{m-n}{b}\pi\right]}{(m-n)^3}
+\frac{\sin\left[\frac{m+n-1}{b}\pi\right]}{(m+n-1)^3}\right)     \\[4mm]
&-\frac{2b}
{\pi^2}\left(\frac{\cos\left[\frac{m-n}{b}\pi\right]}{(m-n)^2} 
+\frac{\cos\left[\frac{m+n-1}{b}\pi\right]}{(m+n-1)^2}\right) \; ,\\[4mm]
\end{array}\right. 
\eeq
and 
$$
H^{(-)}_{m,n}=\int_{-b}^{b}\psi_m^{(-)}{\hat H}\psi_n^{(-)}\,dx =
$$
\beq
=\left\{
\begin{array}{ll}
m=n:
&   \frac{\hbar^2_{ef}}{2}\left(\frac{n}{b}\pi\right)^2+\frac{2}{3b} \\[2mm]
&   -\frac{2b^2}{\pi^3(2n)^3}\sin\left[\frac{2n}{b}\pi\right]        \\[2mm]
&   +\frac{2b}{(2n)^2\pi^2}\cos\left[\frac{2n}{b}\pi\right]\\[5mm] 
m\neq n:
&   \frac{2b^2}{\pi^3}\left(\frac{\sin\left[\frac{m-n}{b}\pi\right]}{(m-n)^3}
    -\frac{\sin\left[\frac{m+n}{b}\pi\right]}{(m+n)^3}\right)  \\[4mm]
&   -\frac{2b}{\pi^2}
    \left(\frac{\cos\left[\frac{m-n}{b}\pi\right]}{(m-n)^2}
    -\frac{\cos\left[\frac{m+n}{b}\pi\right]}{(m+n)^2}\right) \; .\\[4mm]
\end{array}\right. 
\eeq
The exact energy splitting is obtained by numerical 
diagonalization, in quadruple precision,  
of the quantum Hamiltonian. We took a $4800\times 4800$ matrix, 
thereby achieving $31$ valid digits for the lower levels that we consider. 
In figure 1 we plot 
the negative logarithm of $\Delta \bar{E}$ as a function of the 
mean energy $\bar{E}$ of pairs of almost degenerate 
consecutive energy levels. We note a very 
good agreement between the exact and the Landau splittings. 
To resolve the differences, in figure 2 we show  
the tunneling amplitude $A$ 
(= the expression (29) without the exponential tunneling factor) 
as a function of the mean energy $\bar{E}$ of 
pairs of almost degenerate consecutive energy levels. 

\begin{figure}
\begin{center}
\epsfig{file=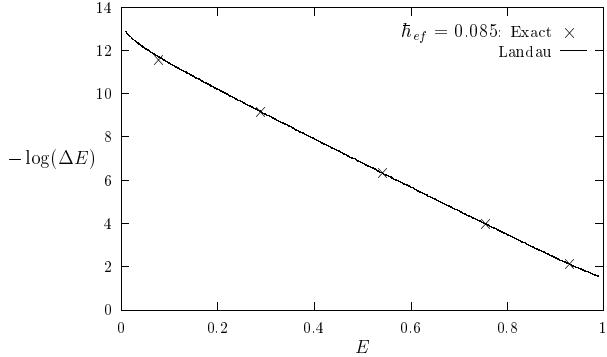, width=8cm}
\caption{Negative logarithm of $\Delta \bar{E}$ {\it vs} mean energy 
$\bar{E}$ of pairs of almost degenerate consecutive energy levels. 
Inverted harmonic oscillator with $\hbar_{eff}=0.085$ and 
$b=1.5$.}
\end{center}
\end{figure}

\begin{figure}
\begin{center}
\epsfig{file=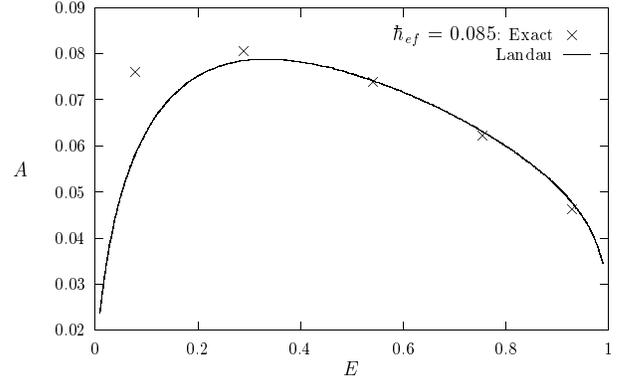, width=8cm}
\caption{Tunneling amplitude $A$ {\it vs} mean energy $\bar{E}$ of 
pairs of almost degenerate consecutive energy levels. 
Inverted harmonic oscillator with $\hbar_{eff}=0.085$ and 
$b=1.5$.}
\end{center}
\end{figure}

\section{Quartic potential}

In this section we consider the well known quartic potential, given by 
\beq
V(x)=-2Bx^2+Ax^4 \; ,
\eeq
where the parameters $A$ and $B$ are related to the potential 
barrier $V_0$ and to the position of the minimum $x_0$ by 
\beq
V_0=\frac{B^2}{A},~~~x_0=\sqrt{\frac{B}{A}} \, .
\eeq
By using the following reduced variables
\beq
\bar{x}=\frac{x}{x_0} \; ,~~~\bar E=\frac{E}{V_0}\, , 
~~~\hbar_{eff}=\frac{\hbar}{x_0\sqrt{mV_0}} \, ,
\eeq
the quantum Hamiltonian operator of the system can be written 
\beq
{\hat H}= -{\hbar_{eff}^2\over 2}{\partial^2 \over \partial \bar{x}^2} 
-2\bar{x}^2+\bar{x}^4 \, . 
\eeq
Let $|\psi\rangle$ be a state of the quartic potential. It can be 
written in terms of the eigenstates $|n\rangle$ of the harmonic 
oscillator
\beq
|\psi\rangle = \sum_{n=0}^{\infty}c_n|n\rangle \, .
\eeq
The Schr\"odinger equation for the harmonic oscillator is 
\beq
-\frac{\hbar^2_{ef}}{2}\frac{\partial^2|n\rangle}{\partial 
x^2} +\frac{\omega^2}{2}x^2 |n\rangle =E_n|n\rangle \, .
\eeq
By introducing the creation and annihilation operators
$$
\hat{a}^+=\frac{1}{\sqrt{2\hbar_{ef}\omega}}(\omega x-i\hat p) \, , 
$$ 
\beq
\hat{a}=\frac{1}{\sqrt{2\hbar_{ef}\omega}}(\omega x+i\hat p) \, ,
\eeq
which have the following properties 
$$
\hat{a}^+|n\rangle=\sqrt{n+1}|n+1\rangle \, ,
$$
\beq
\hat{a}|n\rangle=\sqrt{n}|n-1\rangle \, ,
\eeq
we get the matrix elements of the quantum Hamiltonian of the quartic 
potential 
\beqa
H_{n,m} & = &\langle n|\hat
H|m\rangle  \nonumber \\[2mm]
& = &\delta_{n,m}\left(\hbar_{ef}\left(\frac{\omega}{4}
          -\frac{1}{\omega}\right)(2n+1)\right) \nonumber \nonumber \\[2mm]
&  & + \delta_{n,m}\left(
      \frac{3\hbar_{ef}^2}{4\omega^2}(2n^2+2n+1)\right)  \nonumber  \\[2mm]
&  & + \delta_{n+2,m}\sqrt{(n+1)(n+2)}  \\[2mm]
&   & \;\;\times\left(-\hbar_{ef}
      \left[\frac{\omega}{4}+\frac{1}{\omega}\right]
      +\frac{\hbar^2_{ef}}{2\omega^2}(2n+3)\right)  \nonumber \\[2mm]
&  &  + \delta_{n+4,m} \left(\frac{\hbar^2_{ef}}{4\omega^2}\right)
      \nonumber \\[2mm] 
&  &  \;\; \times \sqrt{(n+4)(n+3)(n+2)(n+1)} \, . \nonumber
\eeqa
We calculate numerically, in quadruple precision ($32$ decimal digits), 
the energy levels of the system in the basis of the harmonic oscillator. 
For numerical purposes we took $\omega =2$ and the dimensionality 
$1000\times 1000$, thereby again achieving $31$ valid digits for the lower 
levels that we consider. 
\par
In this way we can compare the exact 
energy splittings with the semiclassical ones, which are 
obtained by using our splitting formula (29). 
For the quartic potential we have
\beqa
\tilde{\sigma}_0(0) & = & -\sqrt{2}\int_0^a\sqrt{V(x)-E}\,dx \nonumber \\[2mm]
&=&
\frac{2\sqrt{2}}{3}\left[\frac{b^2+E}{b}{\rm F}(k)-b{\rm E}(k)\right] \, ,
\nonumber \\[5mm]
\tilde{\sigma}_2(0) & = & -\frac{1}{24\sqrt{2}}\frac{\partial}{\partial
E}\int_0^a \frac{V'' \, dx}{\sqrt{V(x)-E}} \nonumber \\[2mm]
& = & \frac{1}{24\sqrt{2}a^2b}\left[\frac{(2+3E)b^2+E}{\sqrt{1+E}\,b^4} {\rm
F(k)} \right. \nonumber \\[2mm]
& &\;\;\;\;\;\;\;\;\;\;\;\;\;\;\;\;\;\left. 
-\frac{1+3E}{1+E}{\rm E(k)}\right]  \, ,
\eeqa
\newpage
\beqa
\tilde{\sigma}_4(0) & = & -\frac{1}{1152\sqrt{2}}
\left[\frac{\partial^2}{\partial E^2}\int_0^a\frac{V^{''''} \, dx }
{\sqrt{V(x)-E}}
\right. \nonumber \\[2mm]
&  & \;\;\;\;\;\;\;\;\;\;\;\;\;\;\;\;\;\;
\left. -\frac{7}{5}\frac{\partial^3}{\partial E^3}
\int_0^a\frac{V^{''2} \, dx}{\sqrt{V(x)-E}}
\right]  \nonumber \\[2mm]
& = & -\frac{1}{48\sqrt{2}}\left[Z_1{\rm F(k)}+Z_2{\rm E(k)}\right] \, ,
\nonumber
\eeqa
where 
$$
a=\sqrt{1-\sqrt{1+E}},~~~b=\sqrt{1+\sqrt{1+E}},~~~k=\frac{a^2}{b^2} \, .
$$
Moreover we get
\beqa
\int_a^b\frac{1}{p}\,dx & = & \frac{1}{\sqrt{2}}\frac{{\rm F}(m)}{b} \, , 
\nonumber \\[5mm]
\int_a^b\frac{\sigma_3}{p}\,dx & = &\frac{1}{48\sqrt{2}}\frac{\partial^2} 
{\partial
E^2}\int_a^b\frac{V''}{\sqrt{E-V}}\,dx \nonumber \\[2mm]
& = &-\frac{1}{12\sqrt{2}}\left[K_1{\rm
F}(m)+K_2{\rm E}(m)\right] \, , \nonumber \\[2mm]
& & \\
\int_a^b\frac{\sigma^2_3+\sigma_5}{p}\,dx & =  & \frac{1}{2304\sqrt{2}}
\left[\frac{7}{5}\frac{\partial^4}{\partial E^4}
\int_a^b\frac{V^{''2} \, dx }{\sqrt{E-V}} \right. \nonumber \\[2mm]
& &\;\;\;\;\;\;\;\;\;\;\;\;\;\;\;\left.
-\frac{\partial^3}{\partial E^3}
\int_a^b\frac{V^{''''} \, dx}{\sqrt{E-V}}\right] \nonumber \\[2mm]
& = & \frac{1}{1440\sqrt{2}}\left[Q_1{\rm F}(m)+Q_2{\rm E}(m)\right] \, ,
\nonumber 
\eeqa
where $m=\frac{b^2-a^2}{b^2}$, and 
the functions ${\rm F}(m)$ and ${\rm E}(m)$ are the complete elliptic integrals 
of 1st and 2nd kind, respectively, defined by (Abramowitz and Stegun 1972) 
\beq
F(m) = \int_0^1 {dx\over \sqrt{(1-x^2)(1-mx^2)} } \, ,
\eeq
\beq
E(m) = \int_0^1 {\sqrt{1-mx^2}\over \sqrt{1-x^2} } dx \, .
\eeq
The functions $Z_1$, $Z_2$, $K_1$, 
$K_2$, $Q_1$ and $Q_2$ are defined as follows: 
\newpage
\beqa
Z_1 & = & -\frac{1}{480(1+E)^3a^6b^5} \nonumber \\[2mm]
    &   &\times \left(56b^2+135Eb^2+810E^2b^2-165E^3b^2\right. 
         \nonumber \\[2mm]
    &   &\;\;\;\;\;\; \left. +112+333E-1050E^2-375E^3\right) \, ,
         \nonumber \\[5mm]
Z_2 & = &\frac{224+603E+570E^2-705E^3}{480(1+E)^3a^6b^5} \, , 
         \nonumber \\[5mm]
K_1 & = &-\frac{(1-7E)}{16(1+E)^2 a^2b^3} \, , \nonumber \\[5mm]
K_2 & = &\frac{4+5E+9E^2}{16(1+E)^2a^4b^3} \, , \nonumber \\
\eeqa
\beqa
Q_1 & = & -\frac{1}{128(1+E)^{9/2} a^{6} b^{9}} \nonumber \\[2mm]
    &   &\times \left( b^2\left(672+2263E+2730E^2+9595E^3 \right.\right.
         \nonumber \\[2mm]
    &   & \;\;\;\;\;\;\;\;\;\;\;
          \left.\left. -2008E^4-1136E^5-528E^6
          \right)\right. \nonumber \\[2mm]
    &   & \;\;\;\;\;\left.+\left(
          728E+2454E^2+3755E^3\right.\right. \nonumber \\[2mm]
    &   & \;\;\;\;\;\;\;\;\;\;\; \left.\left.+11008E^4-349E^5+528E^6\right)
          \right)\, , \nonumber \\[5mm]
Q_2 & = & \frac{3(56+209E+289E^2+195E^3-165E^4)}{8(1+E)^4a^8b^7} \, .
          \nonumber
\eeqa
In most of the above manipulations we have used the Mathematica software. 
Please note that the quantities (13) needed in (29) are strongly 
divergent at the turning points, but all the expressions 
in (29) can be made finite by taking partial derivatives 
with respect to $E$ of certain finite expressions as in (54) and (55), 
implying that in our formula (25) then all quantities are 
convergent and finite (c.f. Bender {\em et al} 1977, Robnik and
Salasnich 1997a,b).
\par
In figure 3 we show the tunneling amplitude $A$ as a function 
of the mean energy $\bar{E}$. We compare the exact results 
with the semiclassical ones at 1st (Landau), 3rd and 5th 
order in $\hbar$. We observe that, as expected, there is a 
better agreement by increasing the energy because 
at high energy the classical momentum of the particle is large
and thus the de Broglie wavelength sufficiently small for the
semiclassical methods to be applicable. 
In figure 4 (5) we plot the tunneling amplitude $A$ as a 
function $\hbar_{eff}$ for the first (fourth) pair of almost degenerate 
consecutive energy levels. As shown also in table 1, 
the semiclassical results approach the exact ones by increasing 
the perturbative order in $\hbar$. Note that at the 5th order in $\hbar$ 
the agreement with exact result is up to the 8th digit.

\begin{figure}
\begin{center}
\epsfig{file=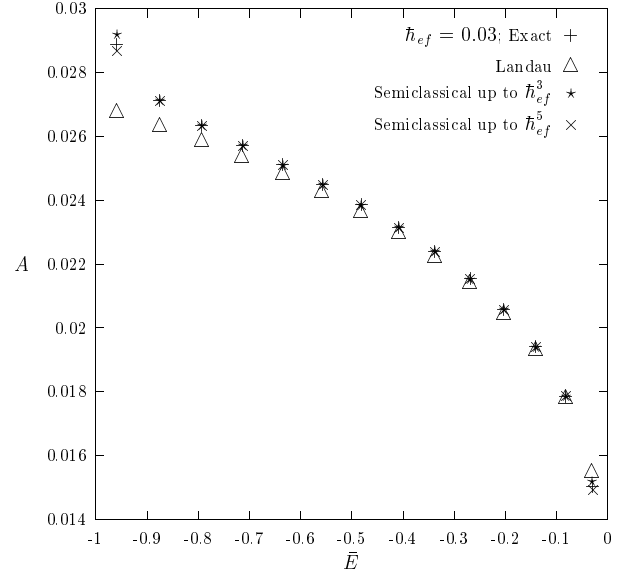, width=8cm}
\caption{Tunneling amplitude $A$ {\it vs} mean energy $\bar{E}$ of 
Quartic potential with $\hbar_{eff}=0.03$.}
\end{center}
\end{figure}

\begin{figure}
\begin{center}
\epsfig{file=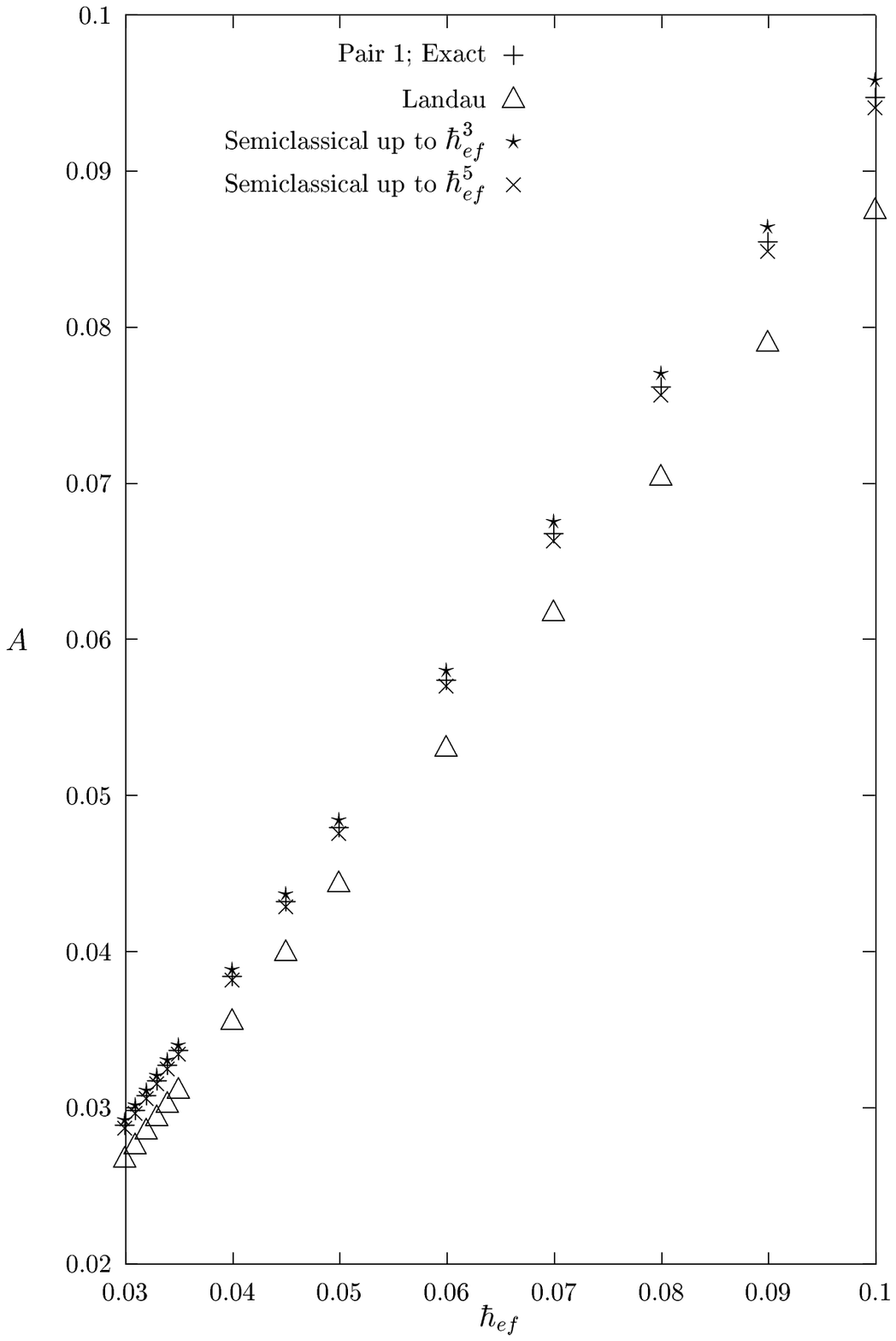, width=8cm}
\caption{Tunneling amplitude $A$ {\it vs} $\hbar_{eff}$ for 
the first pair of almost degenerate consecutive energy levels. 
Quartic potential.}
\end{center}
\end{figure}

\begin{figure}
\begin{center}
\epsfig{file=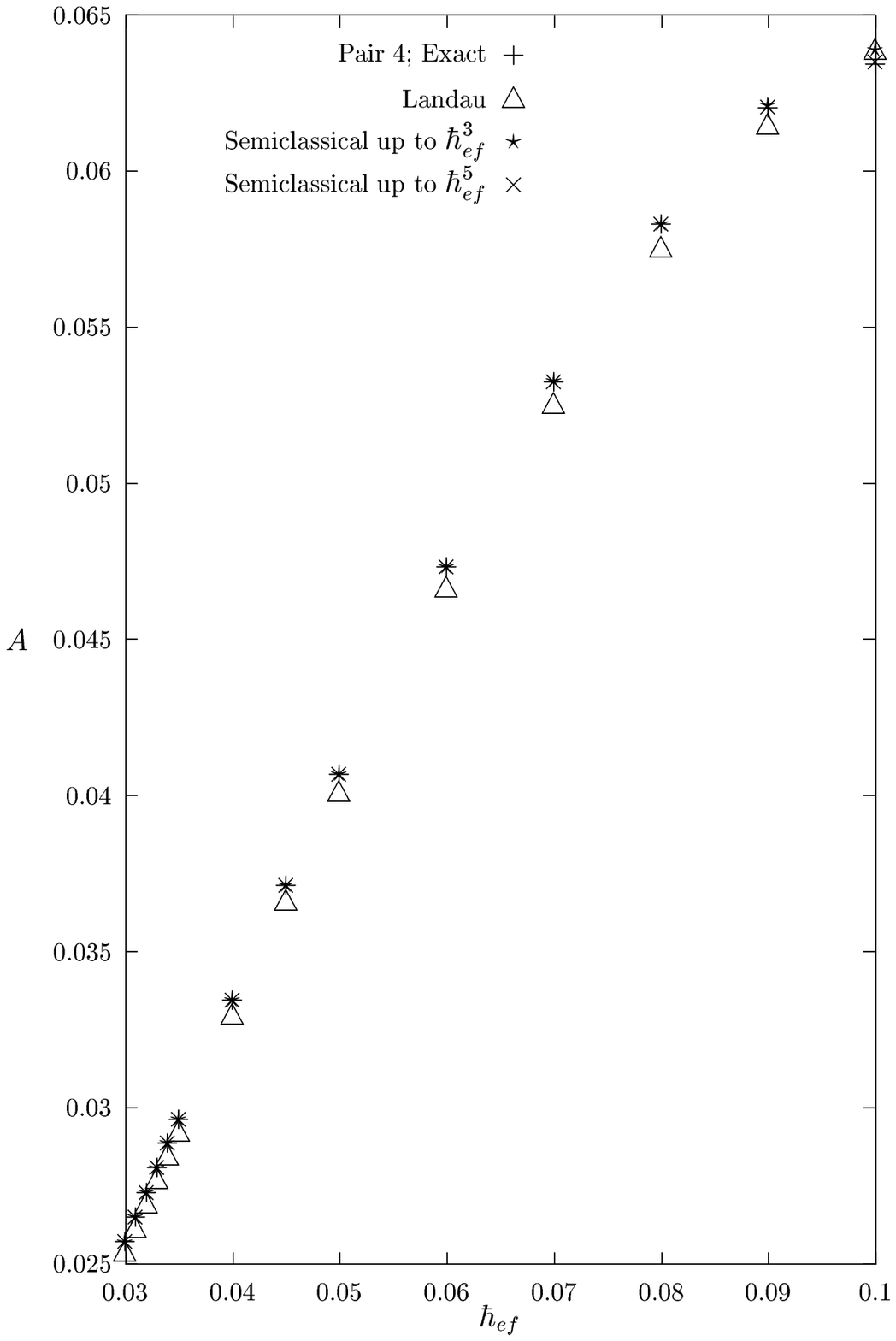, width=8cm}
\caption{Tunneling amplitude $A$ {\it vs} $\hbar_{eff}$ for 
the forth pair of almost degenerate consecutive energy levels. 
Quartic potential.}
\end{center}
\end{figure}

\onecolumn

\begin{table}[ht]
\caption{Energy splitting $\Delta \bar{E}$ for almost degenerate 
pairs at $\hbar_{eff}=0.03$. $\bar{E}_s$ is the exact mean energy, 
$\Delta\bar{E}_{Landau}$, $\Delta\bar{E}_{\hbar^3}$ and 
$\Delta\bar{E}_{\hbar^5} $ are the semiclassical results at 
1st (Landau), 3rd and 5th order, respectively. 
In the last column of the upper table: exact energy splitting 
in units of mean level spacing.}
\begin{center}
\begin{tabular}{l|r|r|r} \hline
 & & & \\[-3mm]
Pair number & $\bar{E}_s$ & $\Delta \bar{E}_{Exact}$ & 
$\Delta \bar{E}_{n}/(\overline{\bar{E}_s^{n+1}-\bar{E}_s^{n})}$ \\ \hline
1  & -0.9578013510838623 & 9.232(2)E-28            &  1.291(6)E-26         \\
2  & -0.8743363182686136 & 6.533698E-25         &  9.140541E-24        \\
3  & -0.7923227236203907 & 2.197647599E-22      &  3.074474586E-21  \\
4  & -0.7118393444104010 & 4.66709587543E-20     & 6.52919406381E-19   \\
5  & -0.6329771419061873 & 7.0097983270055E-18  &  9.8065981172742E-17   \\
6  & -0.5558426330893027 & 7.901944841994295E-16  &  1.10546971247045E-14  \\
7  & -0.4805627052945287 & 6.920784581323967E-14  &  9.68206928062435E-13   \\
8  & -0.4072917698084478 & 4.806630946488390E-12  &  6.72440144371474E-11   \\
9  & -0.3362229365299848 & 2.675674688376366E-10  &  3.74322699989559E-09    \\
10 & -0.2676066496781732 & 1.196767019706533E-08  &  1.67425832453064E-07   \\
11 & -0.2017846855780184 & 4.273586943459416E-07  &  5.97868122857108E-06    \\
12 & -0.1392610433553887 & 1.195177462312803E-05  &  1.67203455862220E-04    \\
13 & -8.088844930248588E-02 & 2.500127371158503E-04  &  3.49763905139656E-03 \\
14 & -2.855578436131249E-02 & 3.352190557610379E-03  &  4.68966210972998E-02   \\

\hline
\end{tabular}
\end{center}
\vspace{5mm}
\begin{center}
\begin{tabular}{r|r|r} \hline
$\Delta \bar{E}_{Landau}(\bar{E}_s)$ & $\Delta \bar{E}_{\hbar^3}(\bar{E}_s)$ & 
$\Delta \bar{E}_{\hbar^5}(\bar{E}_s)$ \\ \hline

8.56386072023299E-28       & 9.{\bf 3}3424576483334E-28 & 9.1{\bf 6}991849898935E-28 \\
6.{\bf 3}4511775769322E-25 & 6.53{\bf 7}96204696428E-25 & 6.533{\bf 3}2080224002E-25 \\
2.1{\bf 5}758453279967E-22 & 2.197{\bf 9}8578660872E-22 & 2.1976{\bf 3}766003663E-22 \\
4.6{\bf 0}423712282166E-20 & 4.667{\bf 3}7049838592E-20 & 4.66709{\bf 2}38092202E-20 \\
6.9{\bf 3}436277656352E-18 & 7.010{\bf 0}0176474235E-18 & 7.00979{\bf 7}08264343E-18 \\
7.8{\bf 3}077639810467E-16 & 7.902{\bf 0}7840692513E-16 & 7.901944{\bf 4}5833250E-16 \\
6.8{\bf 6}703907509728E-14 & 6.9208{\bf 6}233123417E-14 & 6.9207845{\bf 1}430764E-14 \\
4.7{\bf 7}381431395796E-12 & 4.8066{\bf 7}193943203E-12 & 4.8066309{\bf 8}878842E-12 \\
2.6{\bf 5}943820818162E-10 & 2.6756{\bf 9}528644699E-10 & 2.6756747{\bf 7}543535E-10 \\
1.19{\bf 0}31027734624E-08 & 1.1967{\bf 7}788146677E-08 & 1.196767{\bf 1}4995564E-08 \\
4.2{\bf 5}353592752249E-07 & 4.2736{\bf 5}612082292E-07 & 4.27358{\bf 9}20983707E-07 \\
1.19{\bf 0}79278916304E-05 & 1.1952{\bf 4}083756918E-05 & 1.19518{\bf 3}71126262E-05 \\
2.49{\bf 8}35207740361E-04 & 2.50{\bf 1}42635772383E-04 & 2.500{\bf 5}8564705645E-04 \\
3.{\bf 4}6270869995442E-03 & 3.3{\bf 9}489990444765E-03 & 3.3{\bf 3}278741702504E-03 \\

\hline
\end{tabular}
\end{center}
\vspace{5mm}
\end{table}

\twocolumn

\section{Conclusions}
In this work we have taken the first step towards a systematic
improvement of the Landau formula (Landau and Lifshitz 1997),
which is the semiclassical leading order energy level splitting
formula for pairs of almost degenerate levels in double well
potentials. We have developed the algorithm for the semiclassical
$\hbar$ expansion series to all orders for the tunneling
amplitude $A$ (of equation (1)), and thus
calculated explicitly the quantum corrections
up to the 5th order. We have compared the semiclassical 
predictions with the exact results obtained numerically, in 
quadruple precision in case of the quartic double well 
potential. Our approach is based on the usual WKB expansion
in one--dimensional potentials. Thus the 
calculation of higher corrections can in principle be continued by 
the same method, although the structure of higher terms
increases in complexity very quickly. We have also shown
what happens in cases where the assumption implicit in the
Landau formula (namely the linearity of the potential around
the turning points) is not satisfied: We get a different
result even in the leading semiclassical order, and this
has been shown for the double square well potential. 
We should stress that the Landau formula (1997) is indeed quite good 
approximation since it always yields the correct order 
of magnitude (the exponential tunneling factor is always exact) and 
even the tunneling amplitude is correct within the $5$--$50$ \% . 
\par
It is our goal to work out a more direct WKB approach 
to the solution of the multi--minima problem, 
by the contour integration technique, 
based on requiring the single valuedness of the eigenfunction,
as has been done in the case of a single well potential
in (Robnik and Salasnich 1997a,b). This is our future project.
\par
Finally we should mention important applications e.g. in the
domain of molecular physics (Herzberg 1991, Landau and Lifshitz
1997, Cohen-Tannoudji {\em et al} 1993).  
For example, the $NH_3$ molecule can be 
described by a quasi one-dimensional potential $V(x)$ as a function of
the perpendicular distance  $x$ of the nitrogen  $N$ atom from the $H_3$ plane,
and as such it has the double well form, with the top of the
potential barrier at $x=0$. In order to calculate the energy
level splitting of the doublets of vibrational modes one
needs exactly our theory. Another example is the torsional
motion of $C_2H_4$ molecule, where again we encounter an
effectively one-dimensional double well potential. In case
of $C_2H_6$ molecule, we find three potential wells where
the tunneling effects again determine the splittings of  energy
triplet levels and a generalization of our theory would give an 
improved estimate of the splittings.

\section*{Acknowledgments}
This work was supported by the Ministry of Science and
Technology of the Republic of Slovenia and by the Rector's Fund of the 
University of Maribor. L.S. was partially supported 
by the MI41 project of the Italian INFN. 

\section*{Appendix}
The semiclassical splitting formula is obtained by inserting the semiclassical
wavefunction $\psi_0(x)$ for the classically forbidden region $|x|<a$ 
$$
\psi_0 (x)=\frac{C}{\sqrt{|\tilde{p}|}}\exp\left[\frac{1}{\hbar}
\left(\int_a^x
|\tilde{p}|\,dx+\tilde{\sigma}_{even}+\tilde{\sigma}_{odd}\right)\right] \, ;
$$
\beq\label{app}
\tilde{\sigma}_{even}=\sum_{k=1}^{\infty}\hbar^{2k}\tilde{\sigma}_{2k} \, ,
\eeq
$$
\tilde{\sigma}_{2k}(x)=\int_a^x\tilde{\sigma}'_{2k}
(|\tilde{p}(\xi)|)d\,\xi \, ,
$$
$$
\tilde{\sigma}_{odd}=\sum_{k=1}^{\infty}\hbar^{2k+1}\tilde{\sigma}_{2k+1}
\, ,
$$
$$
\tilde{\sigma}_{2k+1}(x)=\int^x\tilde{\sigma}'_k(|\tilde{p}(\xi)|)d\,\xi
\, ,
$$   
into the general expression (8). The potential $V(x)$ can always 
be written in
the way where the maximum of the barrier is at the point $x=0$. 
Considering that,
and inserting (\ref{app}) into (8) we get
\beqa
\Delta E & = & \frac{2\hbar C^2}{m|\tilde{p}|}(|\tilde{p}|
               +\tilde{\sigma}'_{even}
               +\tilde{\sigma}'_{odd})  \nonumber \\[2mm]
         &   & \times
               \exp\left[\frac{2}{\hbar}\left(\int_a^0|\tilde{p}|\,dp
               +\tilde{\sigma}_{even}
               +\tilde{\sigma}_{odd}\right)\right] \, . \nonumber \\[2mm]
\eeqa
The exponent of all terms except the leading one can be expanded in the $\hbar$
power series. The expression valid up to the 5th order in $\hbar$ expansion
reads as
\beqa
\Delta E & = & \frac{2\hbar C^2}{m}\left[\frac{}{}1
               +\hbar 2\tilde{\sigma}_2(0)\right.
               \nonumber \\[2mm]
         &   & +\hbar^2
                \left(\frac{\tilde{\sigma}'_2(0)}{\tilde{\sigma}'_0(0)}
               +2\tilde{\sigma}^2_2(0)+2\tilde{\sigma}_3(0)\right) 
               \nonumber \\[2mm] 
         &   & +\hbar^3
               \left(2\tilde{\sigma}_4(0)+\frac{4}{3}\tilde{\sigma}^3_2(0)+
               4\tilde{\sigma}_2(0)\tilde{\sigma}_3(0)\right)
               \nonumber \\[2mm]
         &   & +\hbar^3\left(\frac{\tilde{\sigma}'_3(0)}
                {\tilde{\sigma}'_0(0)}+
                2\tilde{\sigma}_2\frac{\tilde{\sigma}'_2(0)}
                {\tilde{\sigma}'_0(0)}\right) 
               \nonumber 
\eeqa
\beqa
&  & 
      +\hbar^4
      \left(\frac{\tilde{\sigma}'_4(0)}{\tilde{\sigma}'_0(0)}
      +2\frac{\tilde{\sigma}'_2(0)}{\tilde{\sigma}'_0(0)}
      \left(\tilde{\sigma}^2_2(0)+\tilde{\sigma}_3(0)\right)\right)
      \nonumber \\[2mm]
&  &  +\hbar^4\left(
       2\tilde{\sigma}_2(0)\frac{\tilde{\sigma}'_3(0)}{\tilde{\sigma}'_0(0)} 
       +2\tilde{\sigma}_5(0)+\frac{2}{3}\tilde{\sigma}_2^4(0)\right)
      \nonumber \\[2mm]
&  & + \left.\hbar^4\left(
      2\tilde{\sigma}_3^2(0)+4\tilde{\sigma}_2(0)\tilde{\sigma}_4(0)+
      4\tilde{\sigma}_2^2(0)\tilde{\sigma}_3(0)\frac{}{}\right)\right]
      \nonumber \\[2mm]
&  &  \times \exp\left[-\frac{2}{\hbar}\int_0^a|\tilde{p}|\,dx\right] \, .
      \nonumber 
\eeqa
Using the recursive relation for $\tilde \sigma'_k$ we observe that certain
combinations in the tunneling amplitude in the upper equation are the integrals
of functions identical to 0, evaluated in $x=0$. In the term that stands with
$\hbar^2$ such a combination is 
\beqa
&& \frac{d}{dx}\left(\frac{\tilde{\sigma}'_2(x)}{\tilde{\sigma}'_0(x)}+2\tilde{\sigma}_3(x)\right)=  \nonumber \\[2mm]
&& =\frac{2\tilde{\sigma}'_3(x)\tilde{\sigma}'_0(x)+2\tilde{\sigma}'_2(x)\tilde{\sigma}'_1(x)+\tilde{\sigma}''_2(x)}{\tilde{\sigma}'_0(x)}=0 \, . 
\nonumber \\[2mm]
\eeqa
The same combination multiplied by  $\tilde{\sigma}_2$ and $\tilde{\sigma}_2^2$
can be also found in the brackets after $\hbar^3$ and $\hbar^4$.
In the bracket after $\hbar^4$  one more combination like that can be found
$$
    \frac{d}{dx}\left(\frac{}{}
    2\tilde{\sigma}_5(x)+2\tilde{\sigma}^2_3(x)
    +\frac{\tilde{\sigma}'_4(x)}{\tilde{\sigma}'_0(x)}
    +2\tilde{\sigma}_3(x)\frac{\tilde{\sigma}'_2(x)}
    {\tilde{\sigma}'_0(x)}\right)=
$$
\beqa
    &=&\left(\frac{2\tilde{\sigma}'_5(x)\tilde{\sigma}'_0(x)
    +2\tilde{\sigma}'_4(x)\tilde{\sigma}'_1(x)}{\tilde{\sigma}'_0(x)}\right.
    \nonumber \\[2mm]
    & &\;\;\;+\left.\frac{2\tilde{\sigma}'_3(x)\tilde{\sigma}'_2(x)
    +\tilde{\sigma}''_4(x)}{\tilde{\sigma}'_0(x)}\right)
    \nonumber \\[2mm]
    & &\;+2\tilde{\sigma}_3(x)\frac{2\tilde{\sigma}'_3(x)
    \tilde{\sigma}'_0(x)+2\tilde{\sigma}'_2(x)\tilde{\sigma}'_1(x)
    +\tilde{\sigma}''_2(x)}{\tilde{\sigma}'_0(x)} 
    \nonumber \\[2mm]
    & = & 0 \, . \nonumber \\[2mm]
\eeqa 
In this way the simplified splitting formula valid up to the 5th order
expansion is found \\ \\ 
\beqa
\Delta E & = &\frac{2\hbar C^2}{m} \left[\frac{}{}1+\hbar 2\tilde{\sigma}_2(0)
+\hbar^2 2\tilde{\sigma}^2_2(0)\right.
\nonumber \\[2mm]
&  &
    +\hbar^3
    \left(2\tilde{\sigma}_4(0)+\frac{4}{3}\tilde{\sigma}^3_2(0)+
    \frac{\tilde{\sigma}'_3(0)}{\tilde{\sigma}'_0(0)}\right)
    \nonumber \\[2mm]
& &
    +\hbar^4
    \left.\left(
    2\tilde{\sigma}_2(0)\frac{\tilde{\sigma}'_3(0)}{\tilde{\sigma}'_0(0)}
    \right) \right.
    \nonumber \\[2mm]
& &
    +\hbar^4\left.\left(\frac{2}{3}\tilde{\sigma}_2^4(0)
    +4\tilde{\sigma}_2(0)\tilde{\sigma}_4(0)\right)\right]
    \nonumber \\[2mm]
& & \times\exp\left[-\frac{2}{\hbar}\int_0^a|\tilde{p}|\,dx\right]
\eeqa
and by taking into account the even symmetry of the potential $V(x)$ 
due to which  the first derivatives of all odd $\tilde \sigma_k$ 
evaluated at $x=0$ are zero, we obtain equation (24).

\end{document}